\documentclass{aastex}
\usepackage{apj5}

\shorttitle{43 GHz VLBA Observations of 3C~120}
\shortauthors{G\'omez et al.}

\begin{document}

\title{Monthly 43 GHz VLBA Polarimetric Monitoring of 3C~120 over 16 Epochs:
Evidence for Trailing Shocks in a Relativistic Jet}

\author{Jos\'e-Luis G\'omez\altaffilmark{1}, 
Alan P. Marscher\altaffilmark{2}, 
Antonio Alberdi\altaffilmark{1},
Svetlana G. Jorstad\altaffilmark{2} and Ivan Agudo\altaffilmark{1}
}

\altaffiltext{1}{Instituto de Astrof\'{\i}sica de Andaluc\'{\i}a, CSIC,
Apartado 3004, 18080 Granada, Spain. jlgomez@iaa.es; antxon@iaa.es;
ivan@iaa.es}

\altaffiltext{2}{Institute for Astrophysical Research, Boston University, 725
Commonwealth Avenue, Boston, MA 02215, USA. marscher@bu.edu;
jorstad@rjet.bu.edu}

\begin{abstract}

  We present a 16-month sequence of monthly polarimetric 43 GHz VLBA images of
the radio galaxy 3C~120. The images probe the inner regions of the radio jet
of this relatively nearby superluminal radio galaxy at a linear resolution of
0.07 $h_{65}^{-1}$ pc ($H_o= 65\:h_{65}$ km s$^{-1}$ Mpc$^{-1}$). We follow
the motion of a number of features with apparent velocities between
4.01$\pm$0.08 and $5.82\pm 0.13~h_{65}^{-1}~c$. A new superluminal knot,
moving at $4.29\pm 0.16~h_{65}^{-1}~c$, is observed to be ejected from the
core at a time coincident with the largest flare ever observed for this source
at millimeter wavelengths. Changes in the position angle of this component, as
well as a progressive rotation of its magnetic polarization vector, suggest
the presence of a twisted (resembling a helix in projection) configuration of
the underlying jet magnetic field and jet geometry. We identify several knots
that appear in the wake of the new superluminal component, moving at proper
motions $\sim 4$ times slower than any of the other moving knots observed in
3C~120. These features have properties similar to those of the ``trailing''
shocks seen in relativistic, time-dependent, hydrodynamical and emission
simulations of compact jets. Such trailing compressions are triggered by
pinch-mode jet-body instabilities caused by the propagation of a strong
perturbation, which we associate with the new strong superluminal component.

\end{abstract}


\section{INTRODUCTION}

  The radio galaxy 3C~120 (redshift $z$=0.033) is a powerful and variable
emitter of radiation at radio to X-ray frequencies, probably powered by a
central black hole of at least $3\times10^7 M_{\sun}$ \citep{Ma91,WPM99}. It
was among the first radio jets in which apparent superluminal motion was
detected \citep{Se79,Wa87,Be88,Wa97}. Previous observations using the Very
Long Baseline Array (VLBA) at 22 and 43 GHz \citep{JL98} reveal a very rich
inner jet structure containing up to ten different superluminal
components. Coordinated Millimeter VLBI Array (CMVA) observations at 86 GHz
\citep{JL99}, at an angular resolution of 54 $\mu$as, provide an upper limit
to the size of the core of $\sim 1$ lt-month. Results from long-term
milliarcsecond-scale monitoring at 1.7, 5, and 10.7 GHz \citep{Wa01} allowed
determination of superluminal motions up to at least 150 pc in projection from
the core, as well as evidence for stationary features suggestive of a helical
pattern viewed in projection.

  Further monitoring at a higher frequency, consisting of 16 monthly
polarimetric 22 GHz VLBA observations of 3C~120 \citep{JL00}, explored a more
compact region in the jet, where superluminal components undergo variations in
total and linearly polarized flux densities on timescales of months. [A movie
generated from these 16 total and polarized intensity images can be downloaded
at {\it Science} Online (www.sciencemag.org/feature/data/1052657.shl).] In
this {\it Letter}, we present 43 GHz images corresponding to the same epochs
as the 22 GHz observations presented in \cite{JL00}.

\section{OBSERVATIONS AND DATA ANALYSIS}

  We observed 3C~120 with the 10 antennas of the VLBA at a frequency of 43 GHz
at the following epochs: 1997 Nov. 10, 1997 Dec. 11, 1998 Jan. 11, 1998
Feb. 7, 1998 Mar. 9, 1998 Apr. 10, 1998 May 9, 1998 June 11, 1998 July 11,
1998 Aug. 13, 1998 Sept. 16, 1998 Oct. 26, 1998 Dec. 3, 1999 Jan 10, 1999
Feb. 10, and 1999 Mar. 19. The data were recorded in 1-bit sampling VLBA
format with 32 MHz bandwidth per circular polarization. Reduction of the data
was performed with the AIPS software in the usual manner
\citep[e.g.,][]{Ka95}. Opacity corrections were introduced by solving for
receiver temperature and zenith opacity at each antenna. The instrumental
polarization was determined using the feed-solution algorithm developed by
\cite{Ka95}. The feed D-terms (instrumental polarization) were found to be
very consistent over all sources observed and to remain stable. The absolute
phase offset between right- and left-circularly polarized data (which
determines the polarization position-angle calibration) was obtained by
comparison of the integrated polarization of the VLBA images of several
compact sources (0420-014, OJ287, BL~Lac, and 3C~454.3) with 14 Very Large
Array (VLA) observations at epochs 1997 Nov. 21, 1999 Dec. 14, 1998 Jan. 15,
1998 Feb. 12, 1998 Mar. 7, 1998 Apr. 8, 1998 June 9, 1998 July 11, 1998
Aug. 14, 1998 Sep. 19, 1998 Oct. 29, 1998 Nov. 28, 1999 Feb. 17, and 1999
\includegraphics[scale=0.9]{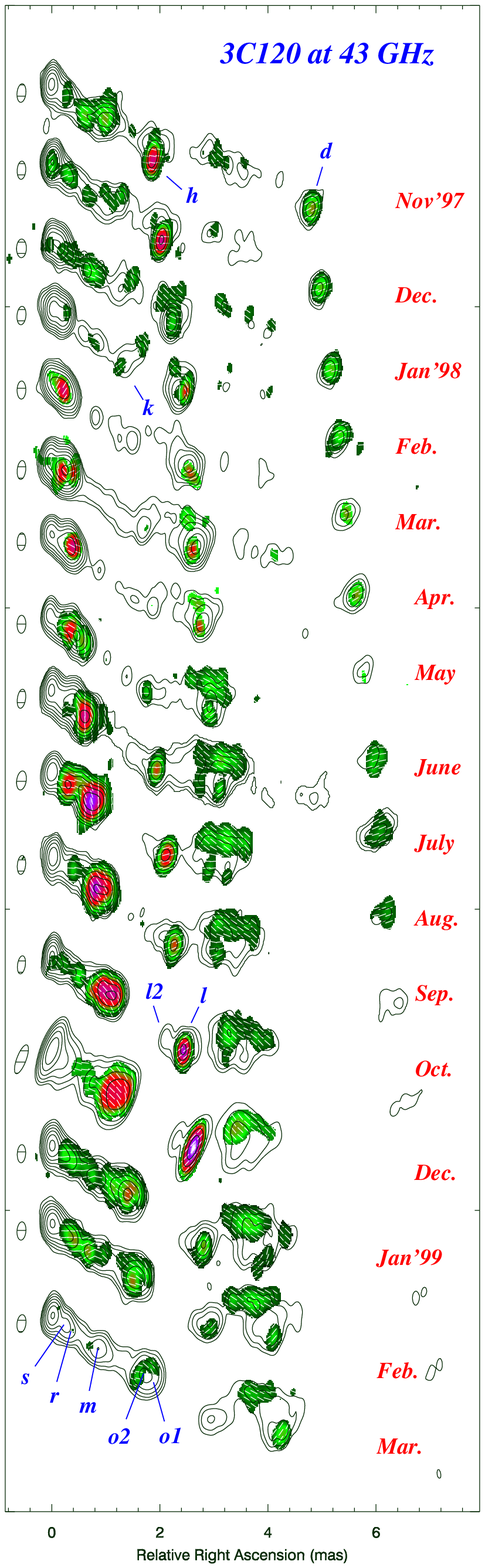}
\figcaption{VLBA images of 3C~120 at 43 GHz. Epochs of observation are
indicated to the right of each image. Vertical image separation is
proportional to the time difference between epochs of observation. Contours
give the total intensity, colors (on a linear scale from green to white) show
the polarized intensity, and bars (of unit length) indicate the direction of
the magnetic polarization vector. Synthesized beams are plotted to the left of
each image, with a typical size of 0.35$\times$0.16 mas. The peak brightness
(r.m.s. noise) in polarization corresponds to 32.4 mJy beam$^{-1}$ (1.4 mJy
beam$^{-1}$). Contour levels for all epochs are in factors of 2 of the bottom
level of 4.4 mJy/beam, except for epochs Feb'98, May'98, Jul'98, Aug'98,
Oct'98, Dec'98, and Jan'99 for which the bottom contour level is 8.8
mJy/beam. For epoch Jun'98 an extra contour at 2.2 mJy/beam has been
plotted. No data was obtained in MK station for epoch Dec'98, which resulted
in a larger synthesized beam. \label{43ghzims}}
\vspace{0.5cm}
\noindent
Mar. 17. Estimated errors in the orientation of the polarization vectors vary
from epoch to epoch, but usually lie in the range of 7--15$^{\circ}$,
confirmed by the stability of the D-terms across epochs (G\'omez et al., in
preparation).

\section{RESULTS}

  The images, plotted in Fig. \ref{43ghzims}, reveal a rich, variable
structure in both total and linearly polarized intensity.  In order to
identify and follow discrete features across epochs, we performed model fits
of the {\it u-v} data with circular Gaussian components using the software
Difmap \citep{Se97}, which was also used to edit the data and make the final
images. Figure \ref{modfits} shows the positions and magnetic polarization
direction for the fitted components at all epochs. Inspection of this figure
reveals that, at a distance of approximately 2 mas from the core (the closest
distance at which we can resolve the jet across its width), bright features
usually lie near the edge of the jet, suggesting limb brightening. This
stratification is even more apparent in polarization: Magnetic vectors on the
northern side of the jet are oblique to the jet axis but constant in
orientation, while those on the southern side rotate as the polarized emission
features move downstream (e.g., components {\it h} and {\it l} in
Fig. \ref{modfits}). This is similar to the behavior observed at 22 GHz
\citep{JL00}.

\subsection{Superluminal Components}

  Because of the complexity and temporal variability of the jet, some of the
fitted components cannot easily be identified across epochs. We concentrate
our discussion on those features that can be followed reliably.  Six of these
components, those labeled in Fig. \ref{43ghzims} as {\it o} (containing
components {\it o1} and {\it o2}), {\it l2}, {\it l}, {\it k}, {\it h}, and
{\it d}, separate from the core at apparent superluminal velocities (proper
motions) of 4.29$\pm$0.16 (1.83$\pm$0.07), 5.38$\pm$0.08 (2.29$\pm$0.04),
5.10$\pm$0.14 (2.17$\pm$0.06), 5.82$\pm$0.13 (2.48$\pm$0.05), 4.12$\pm$0.06
(1.75$\pm$0.03), and 4.01$\pm$0.08 $h_{65}^{-1}\:$c (1.71$\pm$0.03 mas
yr$^{-1}$), respectively.  The separation of these components from the core
versus time is plotted in Fig. \ref{scpm}.

  As observed at 22 GHz \citep{JL00}, components {\it h} and {\it l} brighten
markedly when reaching a distance of about 3 mas from the core (see
Figs. \ref{43ghzims} and \ref{modfits}). G\'omez et al. explained this
brightening, accompanied by a rotation of the magnetic polarization vector, as
interaction between the jet and a cloud with properties intermediate between
those of the broad and narrow emission-line regions.

  The inner jet structure in Fig. \ref{43ghzims} is dominated by the
\includegraphics[scale=0.42]{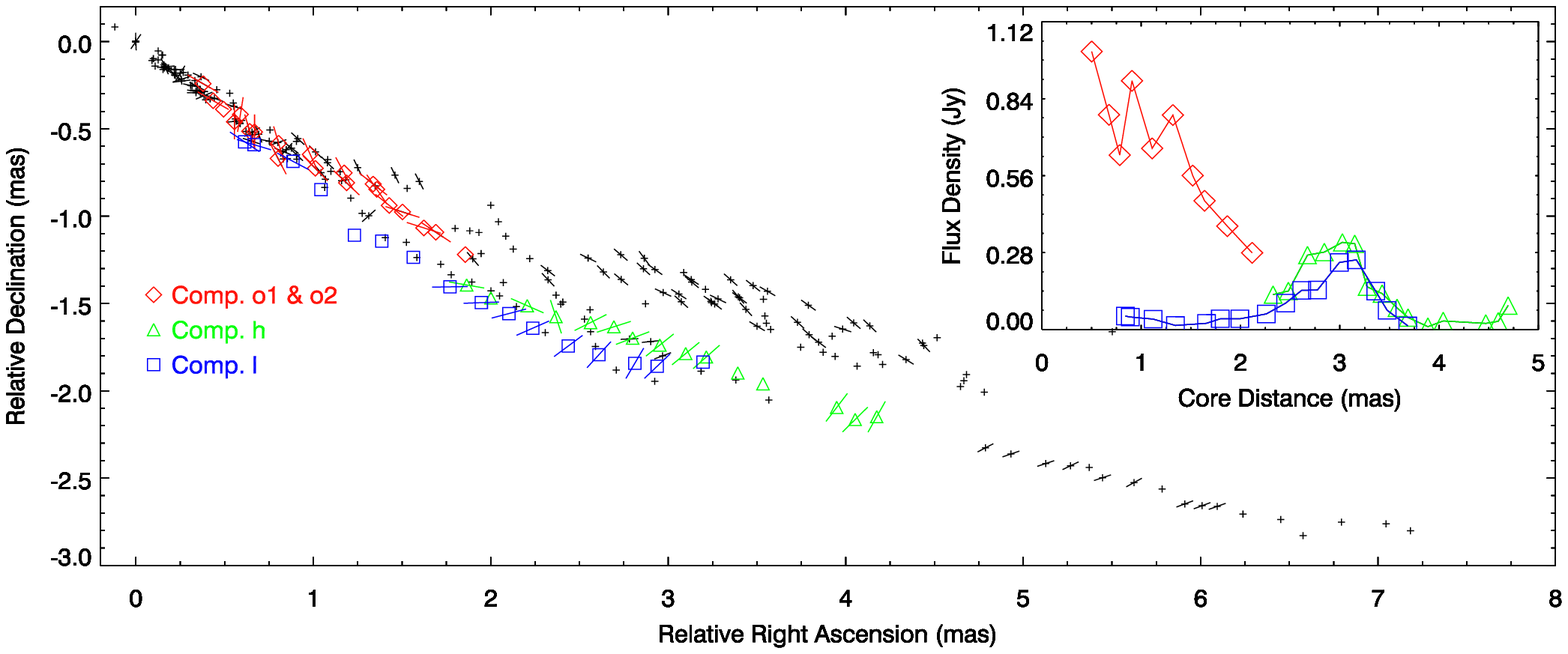}
\figcaption{Positions and magnetic polarization vector orientations of the
components obtained from model fitting of the images shown in
Fig. \ref{43ghzims}. Components {\it o1}, {\it o2}, {\it h}, and {\it l} have
been highlighted with different symbols and colors, with their light curves
plotted in the inset panel. The integrated flux density is plotted at the
position of the brightness centroid for components {\it o1} and {\it o2}.
\label{modfits}}
\vspace{0.5cm}
\noindent
appearance of a new component near the beginning of the largest mm-wave flare
ever observed in 3C~120 (H. Ter\"asranta, private communication). By epoch
1997 December 14 the core had brightened significantly (see Fig. \ref{trlcp}),
after which the new component ({\it o}), appeared downstream of the core
(Fig. \ref{43ghzims}). Figure \ref{trlcp} shows that component {\it o}
presents an extended emission structure that can be split into three different
features (those marked in red in Fig. \ref{trlcp}) between 1998 January 11 and
1998 April 10. These probably do not represent distinct entities, but rather
correspond to complexity in the internal brightness distribution, reminiscent
of the pattern of major disturbances in numerical simulations \citep{JL97}. By
epoch 1998 May 9 the front of knot {\it o} is futher resolved into two
subcomponents, labeled {\it o1} and {\it o2}, with proper motions of
1.87$\pm$0.05 mas yr$^{-1}$ (4.40$\pm$0.12 $h_{65}^{-1}$ c) and 1.78$\pm$0.05
mas yr$^{-1}$ (4.19$\pm$0.11 $h_{65}^{-1}$ c), respectively. The epoch of
ejection (i.e., extrapolated date of coincidence with the core) of component
{\it o1} is 1998.07$\pm$0.03 (see Fig. \ref{trlcp}).

  Although components {\it o1} and {\it o2} move with a relatively constant
proper motion, Fig. \ref{o1o2}a (see also Fig. \ref{modfits}) shows that the
lines between their positions and that of the core vary by 7$^{\circ}$ as they
move between $\sim$0.5 and 2.0 mas from the core. Furthermore, these
variations in position angle are accompanied by rotation of the magnetic
polarization vectors with respect to the local jet axis, as shown in
Fig. \ref{o1o2}b (see also Fig. \ref{modfits}). The initially perpendicular
magnetic vector at epoch 1998 March 9 is observed to align with the jet axis
by 1998 May 8.  This is followed by a rotation of about 60$^{\circ}$ by 1998
June 11. The magnetic vector subsequently rotates more slowly in components
{\it o1} and {\it o2} until it becomes approximately aligned to the jet axis
during the final epochs. Opacity effects, which can produce a rotation of
90$^{\circ}$ \citep[as observed in OJ~287;][]{DG01}, could only be present at
epoch 1998 March 9, when component {\it o} has a nearly flat spectrum
($\alpha\sim-0.06$, $S_{\nu}\propto\nu^{\alpha}$); at the other epochs
components {\it o1} and {\it o2} have steep, optically thin spectra.

  Although Faraday rotation could contribute to the observed rotation of the
magnetic polarization vectors, an alternative interpretation would be the
existence of an underlying helical magnetic field with a progressively
increasing pitch angle until the magnetic field becomes aligned \citep[as
observed at larger scales,][]{Wa87}. This would be in agreement with the
observed variation in the position angles of the core-component separations
\includegraphics[scale=0.47]{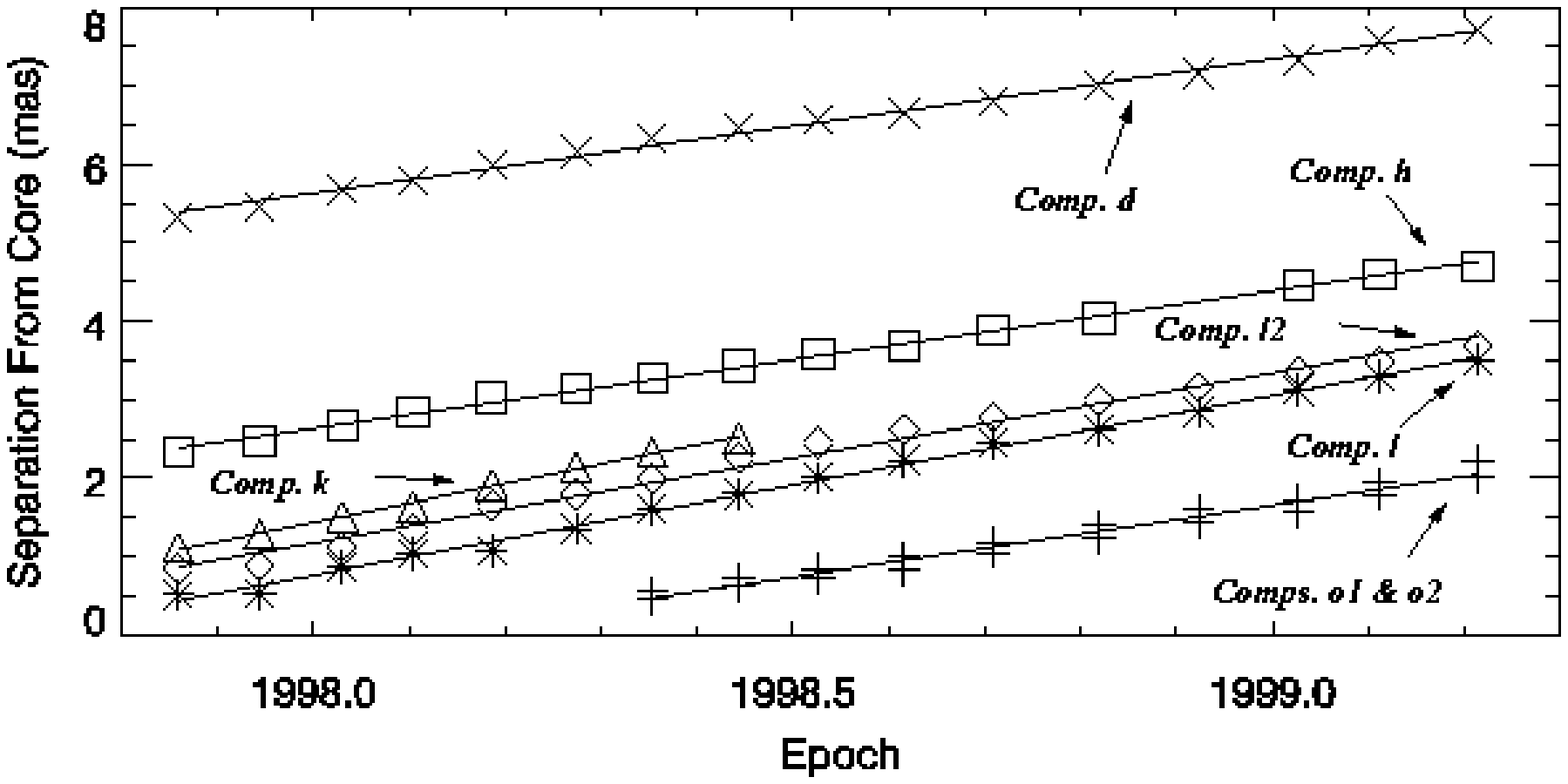}
\figcaption{Projected angular distance from the core as a function of time for
the superluminal components found in 3C~120. Lines show a minimum $\chi^2$
linear fit for each component. A common fit has been used for components {\it
o1} and {\it o2}.\label{scpm}}
\vspace{0.5cm}
\noindent
and the twisted internal structure of component {\it o}, the latter of which
is most clearly visible at epochs 1998 June 11, July 11, and August 13 (see
Fig. \ref{43ghzims}). This interpretation would also agree with the suggested
helical pattern at larger scales \citep{Wa01}.

\subsection{Trailing Components}

  By epoch 1998 May 9 we can distinguish two emission regions in the newly
ejected component {\it o}: the front section (subcomponents {\it o1} and {\it
o2}, plotted in green in Fig. \ref{trlcp}) and the back section ({\it p},
plotted in blue in Fig. \ref{trlcp}). The subsequent evolution of these two
emission regions is significantly different: While {\it o1} and {\it o2} move
with a relatively constant proper motion, Fig. \ref{trlcp} shows that {\it p}
splits into two parts that progressively decelerate and decrease in total flux
more rapidly than do {\it o1} and {\it o2}. Acceleration is also observed to
take place later in component {\it m2}. By epoch 1998 September 16 a similar
split takes place, leading to the appearance of components {\it m} and {\it
m1}. The evolution of the jet following the disturbance that created component
{\it o} therefore involves steady, fast superluminal motion at the front,
followed by the (in some cases temporary) appearance of slower secondary
features in the wake.

  The last epochs in Figs. \ref{43ghzims} and \ref{trlcp} show that the
strongest feature in the wake of component {\it o} is that labeled {\it
m}. This component, which maintains a relatively constant flux density of
$\sim$100 mJy, moves at an apparent speed of 1.16$\pm$0.22 $h_{65}^{-1}\:$c
(0.49$\pm$0.09 mas yr$^{-1}$). This is a factor $\sim 4$ slower than any of
the other moving components detected in 3C~120 (see Fig. \ref{scpm}). Images
at 22 GHz \citep{JL00} also contain component {\it m}, but with a proper
motion compatible with stationarity ($-$0.35$\pm$1.10 mas yr$^{-1}$); however,
the resolution was about twice as coarse as that at 43 GHz. The polarization
of {\it m} is strong at 22 GHz, with magnetic vector direction (relative to
the direction between {\it m} and the core) of 24, 19, and 27$^{\circ}$ and
degree of polarization of 25, 20, and 15\%, at epochs 1999 January 10, 1999
February 10 and 1999 March 19, respectively. Two components even closer to the
core, labeled {\it r} and {\it s}, are also apparent in
Fig. \ref{trlcp}. Their proper motions are the slowest detected in 3C~120:
0.40$\pm$0.03 mas yr$^{-1}$ (0.93$\pm$0.07 $h_{65}^{-1}\:$c) and 0.27$\pm$0.07
mas yr$^{-1}$ (0.63$\pm$0.17 $h_{65}^{-1}\:$c) for {\it r} and {\it s},
respectively.

  The nature of the components that appear on the wake 
of component {\it o} is consistent with the characteristics of the trailing
\includegraphics[scale=0.57]{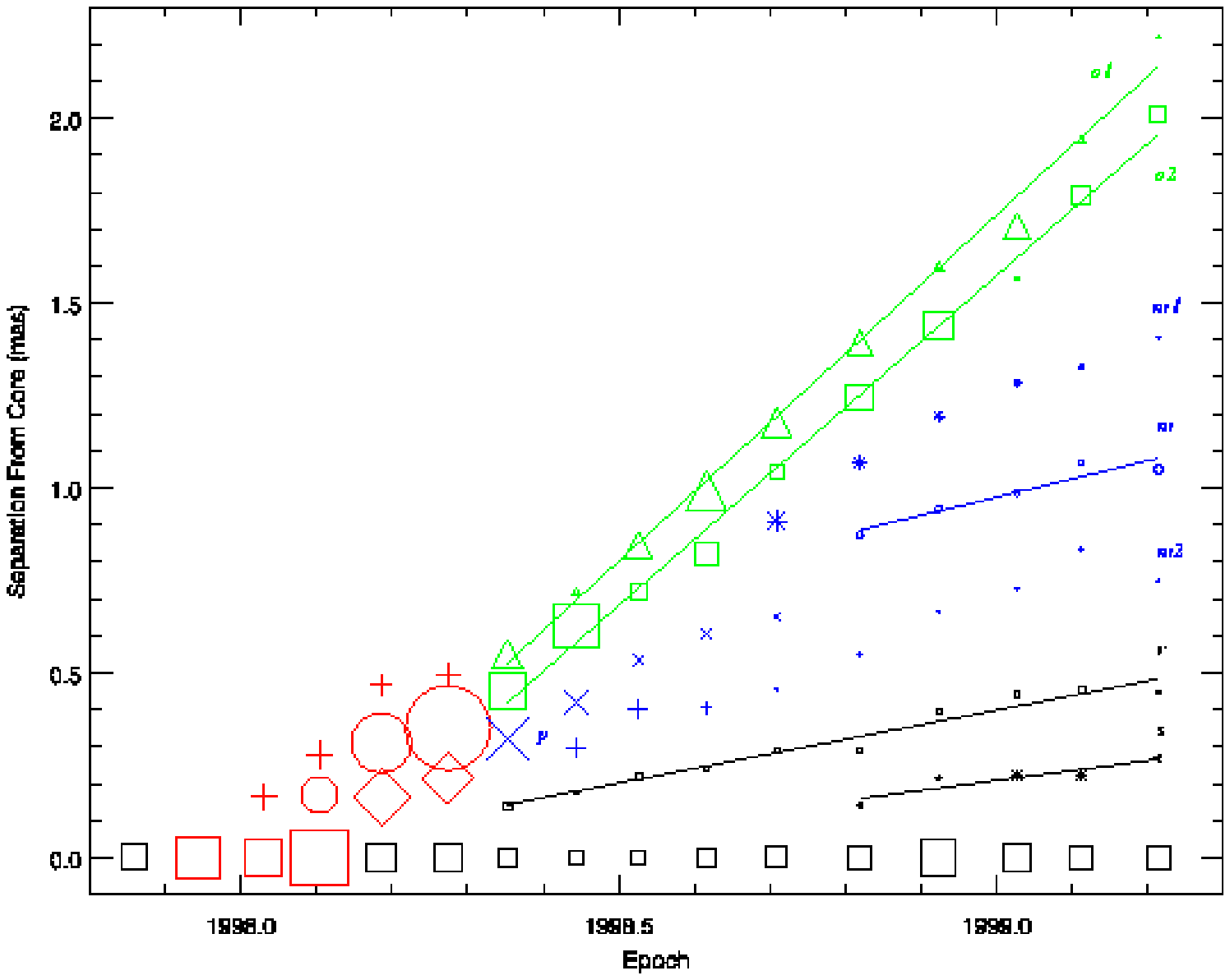}
\figcaption{Projected angular distance from the core vs. time for the jet
features between component {\it o1} and the core. The symbol size is
proportional to the component's total flux density. Those features associated
with the initial evolution of component {\it o} are plotted in red. By epoch
1998 May 9 this component can be resolved into two different sets of
subcomponents, each with a different development: those marked in green (for
components {\it o1} and {\it o2}), and those in blue, corresponding to the
trailing components.\label{trlcp}}
\vspace{0.5cm}
\noindent
features that appear behind a major flow disturbance in relativistic
time-dependent hydrodynamical and emission simulations of jets
\citep{Iv01}. These simulations show that strong jet perturbations (which we
associate with bright superluminal knots) interact with the underlying jet and
external medium as they propagate. This leads to the formation of
recollimation shocks and rarefactions in the wake of
the main perturbation. These formations are triggered by pinch body jet
instabilities. \cite{Iv01} predict that trailing components should appear to
split from the primary component rather than emerge from the core, and to have
significantly slower proper motions than that of the leading strong knot. The
apparent velocities of the trailing features should range from subluminal
closest to the core to more superluminal near the leading knot. This is in
good agreement with the nature of component {\it m}, which indeed appears to
split from component {\it o} and to move at a slower speed. Alternatively,
component {\it m} may correspond to a reverse shock of component {\it o} that
is slower than the forward shock represented by subcomponents {\it o1} and
\includegraphics[scale=0.42]{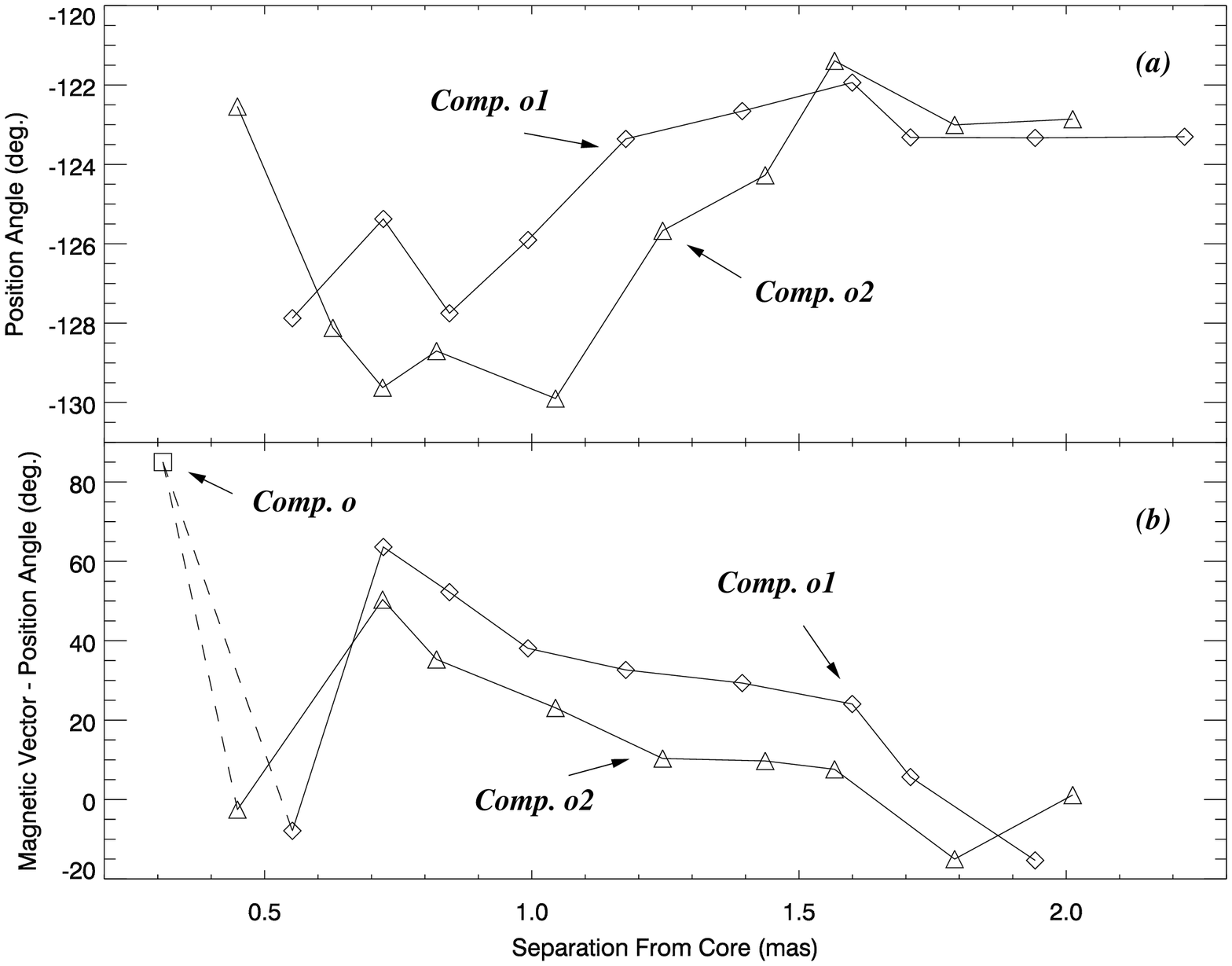}
\figcaption{{\it (a, top)} Position angle and {\it (b, bottom)} orientation of
the magnetic polarization vector relative to the core position angle for
components {\it o}, {\it o1}, and {\it o2} as a function of projected angular
distance from the core.\label{o1o2}}
\vspace{0.5cm}
\noindent
{\it o2} in this interpretation. The subluminal motions of components {\it r}
and {\it s} are also consistent with the predictions for trailing components.

  The good agreement between the changing emission pattern observed in 3C~120
and the structures predicted by 2-D, relativistic, cylindrically symmetric
hydrodynamical and emission simulations \citep{Iv01} points to the value of
such computations for interpreting observations of real jets.
Three-dimensional simulations that are now becoming available promise to
provide even more realistic comparisons.

\begin{acknowledgements}
This research was supported in part by Spain's Direcci\'on General de
investigaci\'on Cient\'{\i}fica y T\'ecnica (DGICYT) grant PB97-1164, by US
National Science Foundation (NSF) grant AST-9802941, and by Fulbright
commission for collaboration between Spain and the United States. The VLBA and
VLA are instruments of the National Radio Astronomy Observatory, a facility of
the NSF operated under cooperative agreement by Associated Universities
Inc. We are grateful to Barry Clark for scheduling \emph{ad hoc} VLA time in
order to determine the polarization position angle calibration.
\end{acknowledgements}

\end{document}